\documentclass{eas}
\usepackage{graphicx}
\begin{document}

\title{Microwave burst with fine spectral structures in a solar flare on 2011 August 9}
\runningtitle{Tan et al: Microwave burst with fine structures}
\author{Baolin Tan, Chengming Tan, \& Yuying Liu}

\address{Key Laboratory of Solar Activity, National Astronomical Observatories of Chinese Academy of Sciences, Datun Road 20A, Chaoyang District, Beijing 100012, China}

\begin{abstract}

On August 9, 2011, there was an X6.9 flare event occurred near the
west limb of solar disk. From the observation obtained by the
spectrometer of the Chinese Solar Broadband Radio Spectrometer in
Huairou (SBRS/Huairou) around the flare, we find that this
powerful flare has only a short-duration microwave burst of about
only 5 minutes, and during the short-duration microwave burst,
there are several kinds of fine structures on the spectrogram.
These fine structures include very short-period pulsations,
millisecond spike bursts, and type III bursts. The most
interesting is that almost all of the pulses of very short-period
pulsation (VSP) are structured by clusters of millisecond
timescales of spike bursts or type III bursts. And there exists
three different kinds of frequency drift rates in the VSP: the
frequency drift rates with absolute value of about 55 - 130 MHz
s$^{-1}$) in the pulse groups, the frequency drift rates with
absolute value of about 2.91 - 16.9 GHz s$^{-1}$) on each
individual pulse, and the frequency drift rates with absolute
value of about 15 - 25 GHz s$^{-1}$) at each individual spike
burst or type III burst.

\end{abstract}
\maketitle
\section{Introduction}

Microwave bursts associated with solar flares offer a number of
unique diagnostic tools to address long-standing questions about
energy release, plasma heating, particle acceleration and
propagation in magnetized plasmas (Rosenberg 1970, Aschwanden
1987, Bastian, Benz, \& Gary, 1998). On 2011 August 9, a most
powerful X6.9 solar flare took place in active region NOAA 11263,
near the west limb on the solar disk. The X6.9 flare event starts
at 08:00 UT, reaches to the maximum at 08:04 UT, and ends at 08:14
UT. It is the largest one in the current solar Schwabe cycle,
resulting in a coronal mass ejection (CME). Accompanied with this
flare, an extremely powerful microwave burst was observed at a
frequency of 2.60 - 3.80 GHz by the Chinese Solar Broadband Radio
Spectrometer in Huairou (SBRS/Huairou) (Fu et al 1995, Fu et al,
2004).

In this work, we will introduce the main features of the microwave
bursts associated with the flare, especially the superfine
structures with millisecond timescale. Section 2 is the
Observations and Analysis, and the conclusions and some
discussions are presented in section 3.

\section{Observations and Analysis}

The upper panel presents the profile of microwave emission at
frequency of 2.84 GHz around the X6.9 flare event during 07:58 -
08:10 UT. The microwave burst starts at 08:01 UT, and ends at
08:06 UT, lasts for only about 5 minutes. As a comparison, we know
that the microwave burst starts at 02:20 UT, ends at 04:10, and
lasts for about 110 minutes in the X3.4 flare event on 2006
December 13 (Tan et al. 2010). The middle panel is the spectrogram
at right-handed circular polarization of a microwave
quasi-periodic pulsation (QPP) observed at 2.60 - 3.80 GHz during
08:03:08 - 08:03:31 UT. This QPP lasts for about 20 s, its average
period is about 0.42 - 0.86 s with averaged value of 0.705 s,
which belongs to very short-period pulsation (VSP) (Tan et al,
2007).

\begin{figure}
 \centering
\begin{tabular}{cc}
\includegraphics[width=1.00\columnwidth]{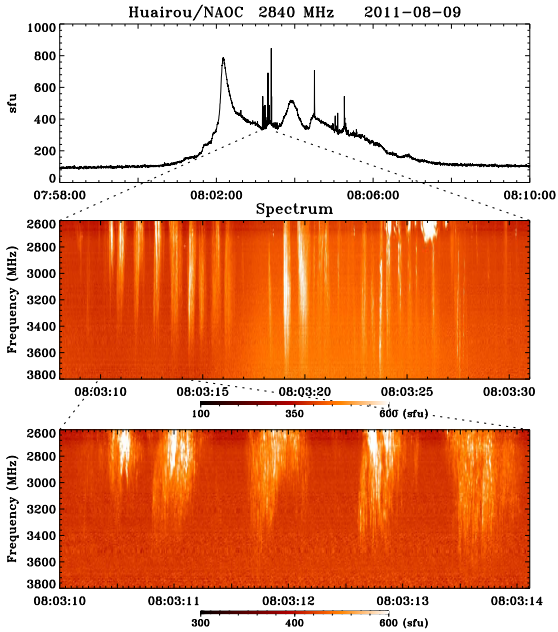}
\end{tabular}
\caption{The upper panel presents the profile of microwave
emission at frequency of 2.84 GHz around the X6.9 flare event
during 07:58 - 08:10 UT. The middle and lower panels are the
spectrograms at right-handed circular polarization of a microwave
quasi-periodic pulsation (QPP) and its expanded spectrogram
observed at 2.60 - 3.80 GHz.} \label{Figure1}
\end{figure}

The spectrogram indicates that the QPP can be divided into three
different pulse groups: the first pulse group begins at 08:03:09
UT, ends at 08:03:17 UT, lasts for 8 s; the second begins at
08:03:18, ends at 08:03:22 UT, lasts for 4 s; and the third begins
at 08:03:23, ends at 08:03:29 UT, and lasts for 6 s. The frequency
bandwidth of each pulses is in the range of 400 - 1100 MHz. The
measurement of the time intervals between each adjacent vertical
bright stripe indicate that the averaged periods are 0.70 s, 0.86
s, and 0.42 s at the first, second, and third pulse group,
respectively. The arranged pattern of bright pulse groups of the
microwave QPP are indicated that there are frequency drifts in
each QPP group (GFDR). The averaged frequency drift rate is 53.3
MHz s$^{-1}$ in the first pulse group, -78.9 MHz s$^{-1}$ in
second pulse group, and from -98.2 MHz s$^{-1}$  turns to 129.5
MHz s$^{-1}$ in the third pulse group. It seems that there are
some dynamic processes occurred in the emission source region.

The lower panel of Figure 1 is a expanded spectrogram at
right-handed circular polarization of several pulses of the QPP
during 08:03:10 - 08:03:14 UT. Here, we may find two main
features: (1) each pulse of the QPP has a frequency drift rate,
and the value is in the range of 2.91 - 16.9 GHz s$^{-1}$,
positive or negative, which is much faster than that in the pulse
groups. (2) Almost all the pulses in the QPP are structured with
groups of millisecond spike bursts, the bandwidth of the spike
bursts is in the range of 20 - 60 MHz, the duration is in the
range of 8 - 24 ms, some spikes have negative frequency drift rate
of 18 - 25 GHz s$^{-1}$.

\begin{figure}
 \centering
\begin{tabular}{cc}
\includegraphics[width=0.45\columnwidth]{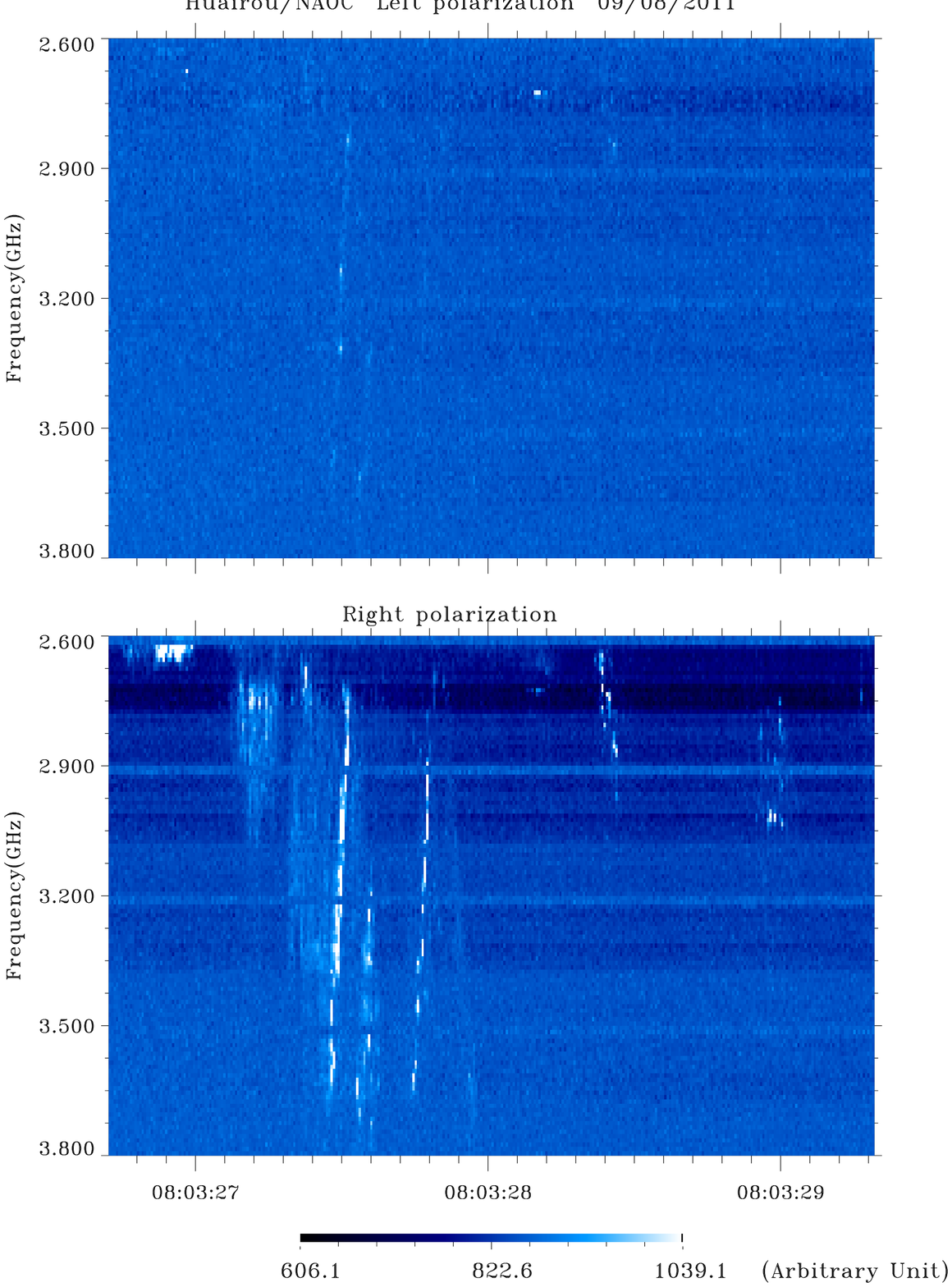}
\includegraphics[width=0.45\columnwidth]{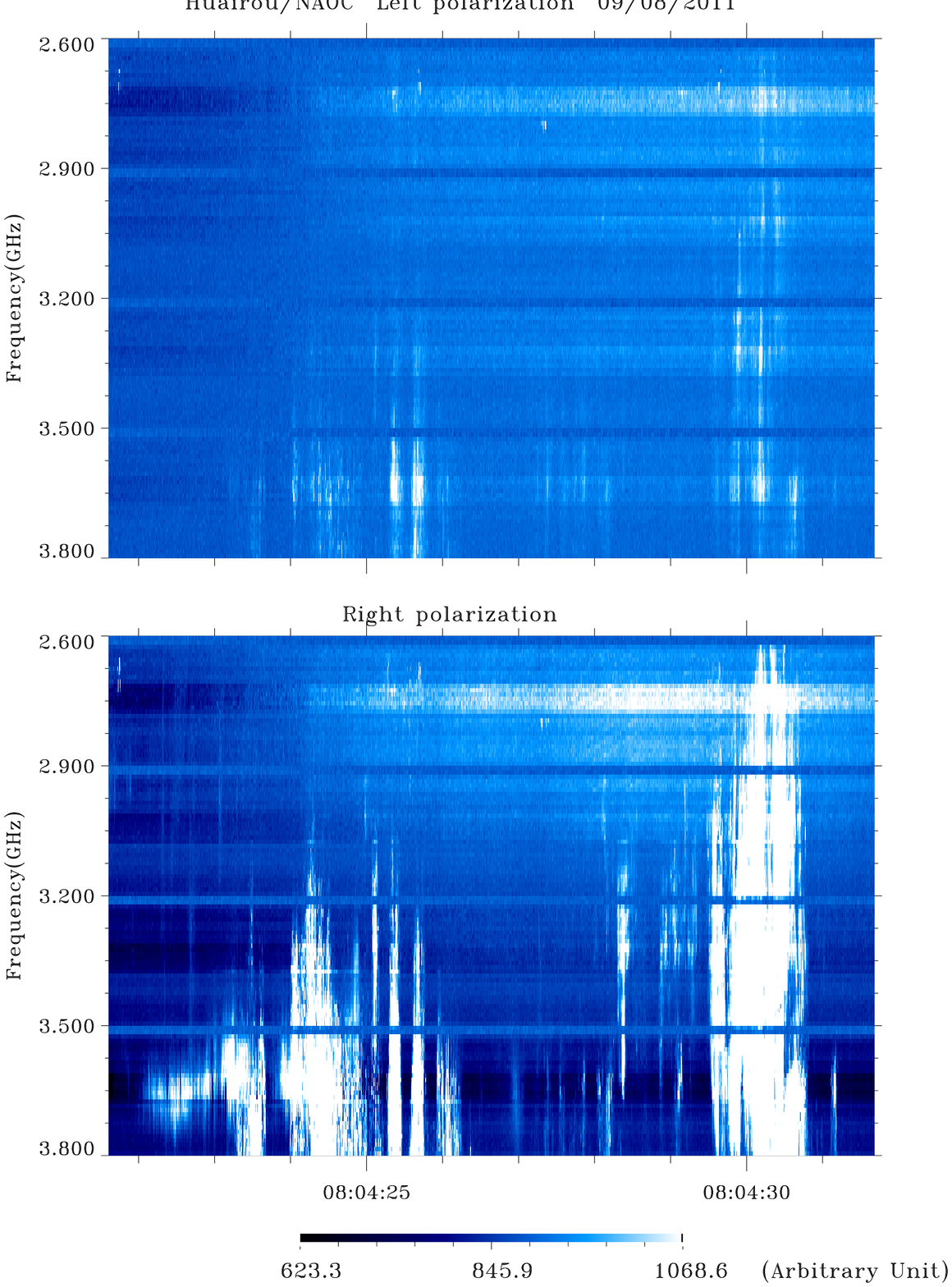}
\end{tabular}
\caption{Left panel is the spectrogram at left- and right-handed
circular polarization of type III bursts. Right panel is the
spectrogram of a series of quasi-periodic pulsations occurred at
high frequency side after the flare peak observed at 2.60 - 3.80
GHz by the Chinese Solar Broadband Radio Spectrometer
(SBRS/Huairou).} \label{Figure1}
\end{figure}

Besides the spike bursts, there are also some narrow band type III
bursts (left panel of Figure 2), which does also form a pulse of
the above QPP. The most interesting is that the frequency drift
rates of the type III bursts in the pulse at 08:03:27.2 -
08:03:28.0 UT is opposite to that at 08:03:28.4 UT. The type III
bursts have the instantaneous duration is about 8 - 24 ms,
frequency bandwidth is about 500 - 1000 MHz, the frequency drift
rate is in the range of from 14.58 GHz s$^{-1}$ to 20 GHz
s$^{-1}$.

The right panel of Figure 2 shows that there are many other QPPs
with different periods occurred in the flare decay phase and in
the high frequency side. Some QPPs have the periods low to several
tens of milliseconds. Some QPPs are narrow bandwidth with only 100
-200 MHz (08:04:22 - 08:04:23 UT, 3.55 - 3.75 GHz, right-handed
circular polarization), there are also some QPPs are broad
bandwidth with about 700 MHz (08:04:23 - 08:04:26 UT, 3.10 - 3.80
GHz, may beyond 3.80 GHz) and 1.10 GHz (08:04:29 - 08:04:31 UT,
2.65 - 3.80 GHz, may beyond 3.80 GHz).

\begin{figure}
 \centering
\begin{tabular}{cc}
\includegraphics[width=0.45\columnwidth]{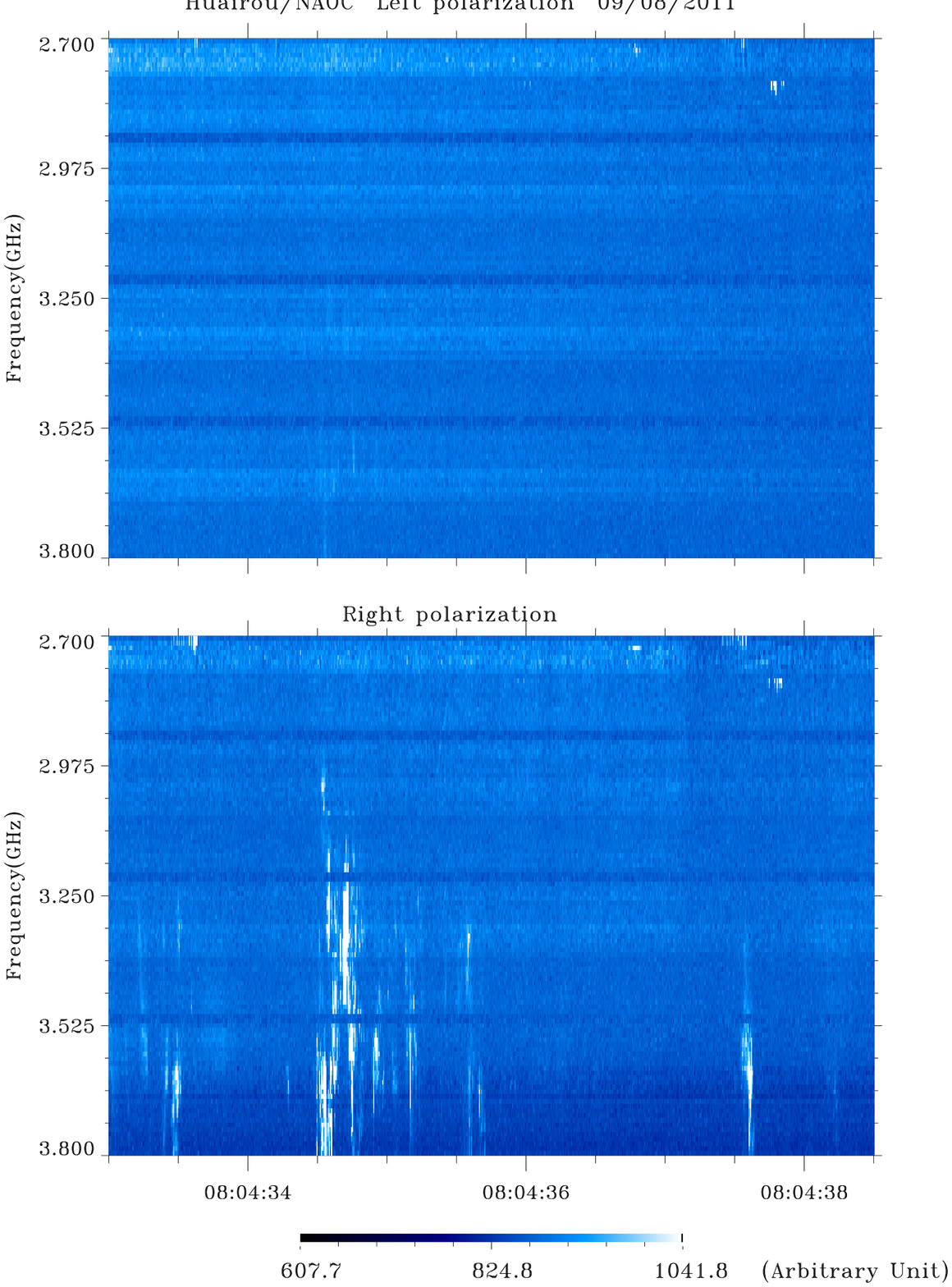}
\includegraphics[width=0.45\columnwidth]{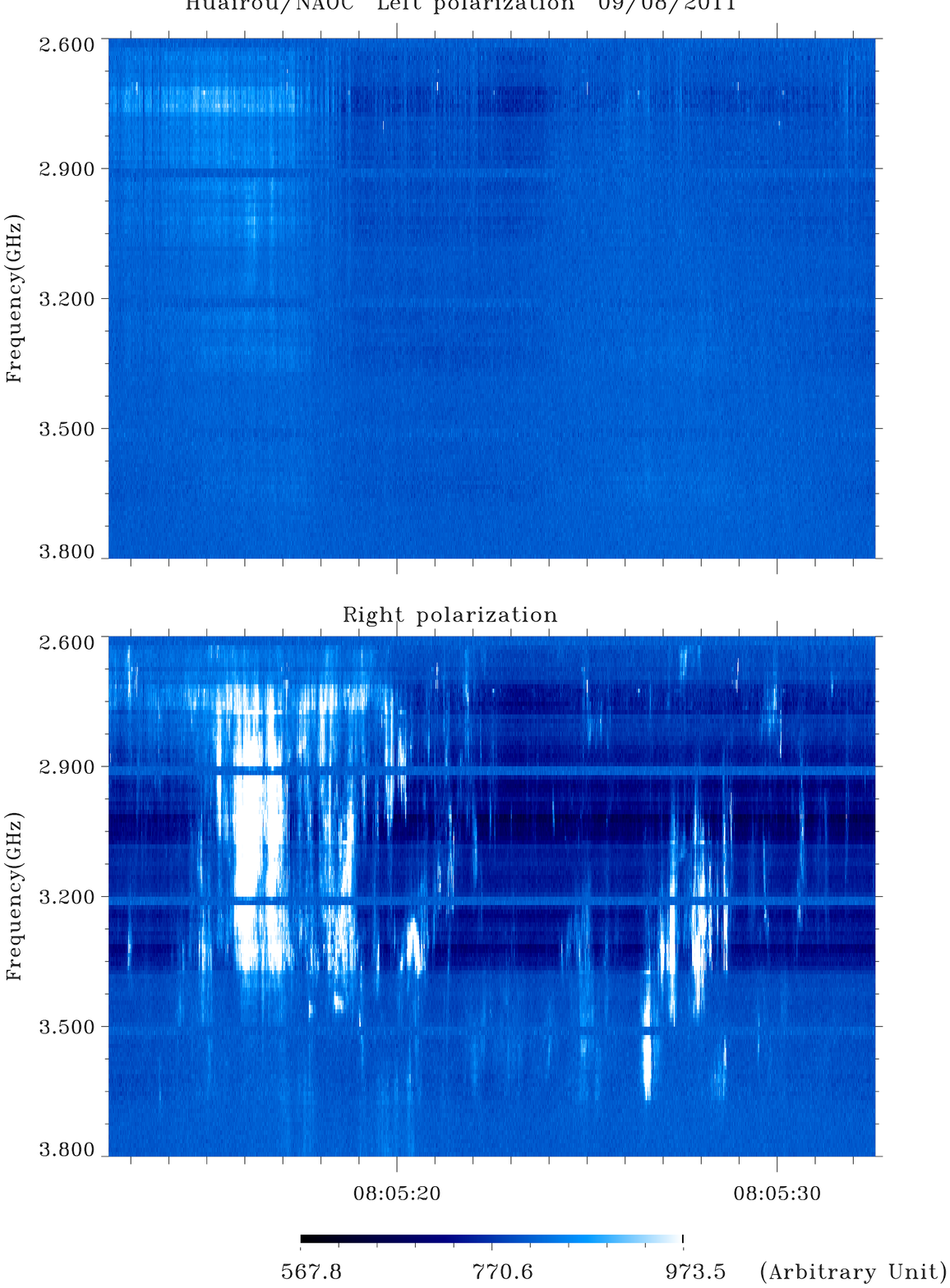}
\end{tabular}
\caption{The spectrogram at left- and right-handed circular
polarization of the microwave spike bursts observed at 2.60 - 3.80
GHz by the Chinese Solar Broadband Radio Spectrometer
(SBRS/Huairou) in the decay phase of the flare event.}
\label{Figure1}
\end{figure}

Additionally, there are some spike bursts in the flare decay phase
sporadically (Figure 3), and some spikes distributed as drifting
pulsating structures (e.g. the structure appeared at 08:05:26 -
08:05:30 UT at frequency of 2.90 - 3.60 GHz, and the frequency
drift rate is about - 150 MHz s$^{-1}$).

\section{Conclusions}

From the above analysis of the observations associated with the
powerful X6.9 flare event on 2011 August 9, we may get the
following primary conclusions:

(1) Dislike the previous observations of X-class flares which
always have long-duration microwave bursts, the above X6.9 flare
event has only a short-duration microwave burst of about only 5
minutes.

(2) During the short-duration microwave burst, there are several
kinds of fine structures on the spectrogram. These fine structures
include very short-period pulsations, millisecond spike bursts,
and type III bursts. The most interesting is that almost all of
the pulses of very short-period pulsation are structured by
clusters of millisecond timescales of spike bursts or type III
bursts.

(3) There exists three different kinds of frequency drift rates in
the VSP: the frequency drift rates with absolute value of about 55
- 130 MHz s$^{-1}$) in the pulse groups, the frequency drift rates
with absolute value of about 2.91 - 16.9 GHz s$^{-1}$) on each
individual pulse, and the frequency drift rates with absolute
value of about 15 - 25 GHz s$^{-1}$) at each individual spike
burst or type III burst.

The flare-associated microwave QPP can provide information on
solar flaring regions and give the possible insight into coronal
plasma dynamics, such as to remote the microphysics of primary
energy releasing regions (Nakariakov \& Milnikov 2009). The
different magnitude of frequency drift rate may reflect the
different kinematics of the microwave emission source regions
(Kliem et al 2000).

~

\textbf{Acknowledgments} This work is mainly supported by NSFC
Grant No. 11103044, 10921303, MOST Grant No. 2011CB811401, the
National Major Scientific Equipment R\&D Project ZDYZ2009-3.


\end{document}